\documentstyle[aps,preprint,prl]{revtex}
\input{epsf}
\begin{document}
\draft
\preprint{To be published in Phys.\ Rev.\ Lett.}
\title{Electronic structure of intentionally disordered AlAs/GaAs
superlattices}
\author{Kurt A. M\"ader, Lin-Wang Wang, and Alex Zunger}
\address{National Renewable Energy Laboratory, Golden, CO 80401}
\date{Received: 5 October 1994}
\maketitle
%{\let\clearpage\relax
%\twocolumn[%
%\widetext\leftskip=0.10753\textwidth \rightskip\leftskip
%
\begin{abstract}
We use realistic pseudopotentials and a plane-wave basis to study the
electronic structure of non-periodic, three-dimensional, 2000-atom
(AlAs)$_{n}$/(GaAs)$_{m}$ (001) superlattices, where the individual
layer thicknesses $n,m \in \{1,2,3\}$ are randomly selected.  We find
that while the band gap of the equivalent ($n = m = 2$)  {\em
ordered\/} superlattice is indirect, random fluctuations in layer
thicknesses lead to a {\em direct\/} gap in the planar Brillouin zone,
strong wavefunction localization along the growth direction,  short
radiative lifetimes, and a significant band-gap reduction, in agreement
with
experiments on such intentionally grown disordered superlattices.
\end{abstract}
%71.50.+t Localized single-particle electronic states (excluding impurities)
%73.20.Dx Electron states in low-dimensional structures (including quantum
%         wells, superlattices, layer structures, and intercalation compounds)
%78.55.-m Photoluminescence
%78.66.-w Optical properties of thin films, surfaces, and layer structures
%         (superlattices, heterojunctions, and multilayers)
%78.66.Fd III-V semiconductors
%
%]}
% preprint:
%\clearpage

% body of paper here

\narrowtext

Ordered semiconductor superlattices---produced routinely by epitaxial
crystal growth techniques---are widely recognized for their unique
electronic and optical properties \cite{SL}.  To tailor the electronic
properties (e.g., through band folding) one aims at growing ordered
superlattices (o-SL) with {\em definite values\/} of the thicknesses
$n$ and $m$ in $(A)_{n}/(G)_{m}$.  One could, however, deliberately
grow a {\em
disordered\/} superlattice (d-SL) \cite{Chomette86,Sasaki89}, where
the individual layer thicknesses $n,m,n',m',n'',m'',\dots$ are
selected at random according to a given probability distribution
$p_{\alpha}(n)$ ($\alpha=A,G$).
While the electronic structure of an o-SL is characterized by
extended states and the formation
of mini-bands, a truly one-dimensional disordered
system (described, e.g., by the Anderson model) shows
carrier localization and absence of dispersion \cite{Borland63,Papa76}.
Sasaki et al.\ \cite{Sasaki89} grew
$\sim$1000 monolayer (ML) thick AlAs/GaAs d-SL's (i.e., $A =$ AlAs, $G
=$ GaAs) with $n$ and $m$ chosen from a set of small integers
$\{1,2,3\}$, viz., $p_{A} = p_{G} = p$, with
$p(1)=p(2)=p(3)=\frac{1}{3}$.  Since the average layer thickness is 2
monolayers, one can think of this d-SL as evolving from an
$(A)_{2}/(G)_{2}$ o-SL by random substitution of $A_{2}$ and $G_{2}$
layers by $A_{1}$, $A_{3}$, $G_{1}$, and $G_{3}$ layers
(``$\delta$-doping'').  Such disordered superlattices have shown
surprising and unique optical properties relative to their parent o-SL
\cite{Sasaki89}: (a) strong and initially fast decaying (lifetime
$\tau = 0.25$ ns at $T =$ 77 K) photoluminescence (PL) intensity even
though the equivalent o-SL has an indirect band gap and thus emits
both weakly and slowly, (b) a large
red shift ($\sim$60 meV) of the PL peak with respect to the equivalent
o-SL, and (c) an order of magnitude slower rate of
reduction of the PL intensity with temperature.  These unusual
properties of d-SL's appear very attractive for optoelectronic
applications \cite{Sasaki89}.

In modelling the electronic structure of a d-SL
\cite{Chomette86,Pavesi89,EGWang94}, one faces difficulties arising
from the existence of two entirely different length scales: (i) The
lack of translational symmetry requires the use of unit cells with a
macroscopic length $N \approx 1000$ ML, equal to the {\em total
length\/} of the d-SL ($N d \approx 300$ nm, where $d$ is the
monolayer thickness). (ii) The spatial variations of the electron
potential, however, occur on a microscopic length scale of about $0.1$
nm.  While it is possible to rescale the microscopic length scale by
replacing the periodic atomic potential by an external, rectangular
potential \cite{Chomette86,Pavesi89}, this approach fails to describe
the band structure (e.g., the indirect gap of the
(AlAs)$_{2}$/(GaAs)$_{2}$ o-SL) in the present regime of rapid layer
fluctuations.
To overcome the problems arising from the existence of two disparate
length scales, we extended a microscopic pseudopotential description
of the electron structure to a macroscopic length scale necessitated
by the absence of translational symmetry.  We use {\em fixed\/}
(screened) atomic pseudopotentials that were carefully fitted to bulk
band structures, effective masses, deformation potentials, band
offsets, and energy levels in superlattices \cite{Mader94}.
The wavefunctions are expanded in
about 30 plane waves per atom.  For an $N$-monolayer
superlattice along (001) with two atoms per monolayer the
corresponding matrix dimension is therefore about
$60N\times60N$. Standard techniques to solve the Schr\"odinger
equation require orthogonalization of the states of interest (i.e., near the
band gap) to all lower-lying states. This leads to an $N^{3}$
scaling of the effort and becomes impractical for structures with
$N\approx100$ ML.  We use instead the ``folded-spectrum'' method
\cite{Wang94}, where eigenstates are obtained directly in an energy
window of interest (e.g., near the band gap), without having to solve
for any of the $\sim 8N$ lower-lying eigenstates first, thus
circumventing the need for orthogonalization.  The effort scales
linearly with $N$, allowing us to use the realistic, three-dimensional
pseudopotentials, and to solve the Schr\"odinger equation in a highly
flexible plane-wave basis even for $N = 1000$ ML.

Application to d-SL's of AlAs/GaAs leads to an
explanation of the large red shift, the enhanced oscillator
strength, and the weak temperature dependence
in terms of band-edge wavefunction
localization.  The source of localization is interesting: In truly
one-dimensional disordered chains {\em all\/} states are in general
localized \cite{Borland63,Papa76}.  However, the laboratory-grown
\cite{Sasaki89}
d-SL's have extended layers in the $(x,y)$ plane (perpendicular to the
disorder direction $z = [001]$), so these {\em quasi\/}
one-dimensional systems retain in fact a three-dimensional character,
and the states need not be localized by disorder.
We will show that the localization in the d-SL originates
mostly from the formation of impurity-like,
localized bound states due to insertion of $\delta$-layers into the
o-SL.
%
%Another source of
%localization is possible, namely the formation of impurity-like,
%localized bound states due to insertion of $\delta$-layers into the
%o-SL. We find that the latter accounts for most of the localization in
%the d-SL for small to moderate degrees of disorder.

%
% fig. 1
%

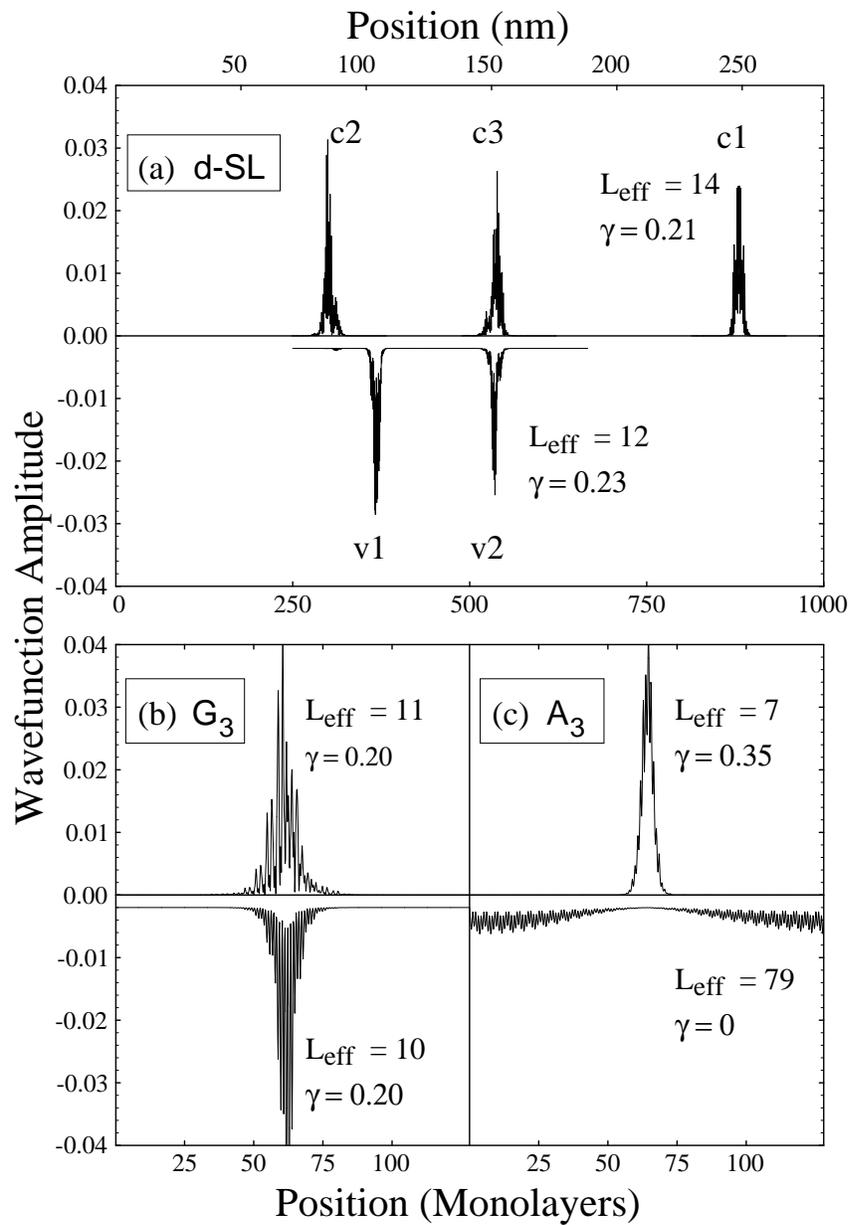
\begin{figure}
%\hbox to \hsize{\epsfxsize=1.00\hsize\hfil\epsfbox{figures/dSL-psi.eps}\hfil}
%\nobreak\bigskip
\caption{Planar average of the wavefunctions squared in
(a) a 1000-mono\-layer d-SL, (b) an o-SL host containing a single
(GaAs)$_{3}$ $\delta$-impurity and (c) a single (AlAs)$_{3}$
$\delta$-impurity.  Unoccupied states are labeled c1, c2,\dots;
occupied states are labeled v1, v2,\dots, and are plotted in the
negative direction with a small offset for clarity.
$L_{\protect\text{eff}}$ and $\gamma$
are given in ML and ML$^{-1}$, respectively.  }
\label{fig:psi}
\end{figure}

In Fig.~\ref{fig:psi}(a) we show the planar wavefunction average
$|\bar\psi_{E}(z)|^{2} = \int d^{2}r_{\perp} |\psi_{E}(\bbox{r})|^2$
of a few occupied and unoccupied band-edge states of a d-SL with a
1000-monolayer unit cell.  We see that these band-edge wavefunctions are
sharply localized. We can quantify the
effective localization length (in monolayer units) for wavefunction
$\psi_{E}$ at energy $E$ as \cite{Papa76}
\begin{equation}
\label{eq:Leff}
	L_{\text{eff}}(E) = \frac{1}{h} \left(\sum_{i}
	|\langle i|\bar\psi_{E}\rangle|^4 \right)^{-1},
\end{equation}
where the sum extends over the grid points $i$ along $z$, and $h$ is
the number of grid points per monolayer.  For a truly extended state,
$L_{\text{eff}}$ is of the order of $N$, while for a state localized
on one site, $L_{\text{eff}}=1$.  We find that $L_{\text{eff}}
\lesssim 15$ ML for the band-edge states.  The asymptotic decay length
$\gamma^{-1}$ can be quantified by
$\langle |\bar\psi_{E}(z)|\rangle \propto e^{-\gamma z}$,
where the angular brackets denote averaging over the fast oscillations
of $|\bar\psi_{E}(z)|$ along $z$.
Near the band edges, the calculated values are $\gamma\approx$ 0.2 ML$^{-1}$.

We now investigate two possible mechanisms for the wavefunction
localization apparent in Fig.~\ref{fig:psi}(a): (i) chemical binding
to $\delta$-like ``impurity layers'' in an otherwise ordered
$(A)_{2}/(G)_{2}$ host, and (ii) a continuous increase of localization
(measured by $\gamma$ or by $L_{\text{eff}}^{-1}$) with increasing
degree of disorder. Scenario (i) is motivated by the fact that in one
dimension an attractive $\delta$-potential always has one bound state,
whereas scenario (ii) is valid for the Anderson model with on-site
disorder obeying a continuous probability distribution
\cite{Papa76}. In our case of a discrete probability distribution
$p(n)$, we connect the two pictures by starting with the reference
o-SL $(A)_{2}/(G)_{2}$, and gradually substituting $A_{2}$ and $G_{2}$
layers by ``wrong'' layers with $n=1$ or $n=3$. We measure the degree
of disorder by counting the relative frequency $R$ of these layers,
i.e.,
\begin{equation}
\label{eq:R}
R = p(1)/p(2) = p(3)/p(2).
\end{equation}
The fully disordered SL has $R=1$, the single $\delta$-layer in an
ordered $(A)_{2}/(G)_{2}$ superlattice corresponds to $R \approx N^{-1}$,
while the perfect o-SL has $R=0$.

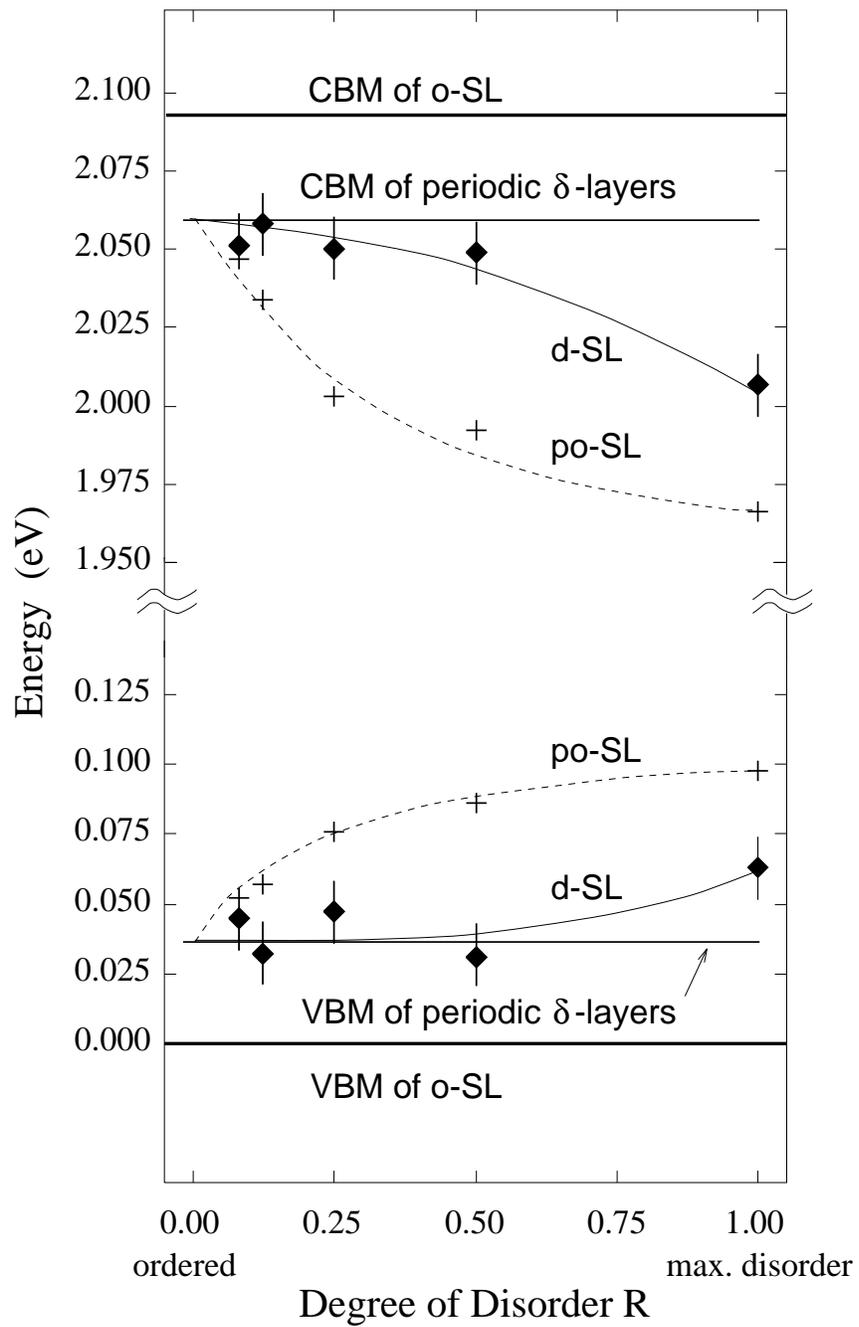
\begin{figure}
%\hbox to \hsize{\epsfxsize=1.00\hsize\hfil\epsfbox{figures/dSL-gaps.eps}\hfil}
%\nobreak\bigskip
\caption{Band-edge energies of disordered (d-SL, diamonds), partially ordered
superlattices (po-SL, pluses), and a periodic array of
$\delta$-layers embedded in an o-SL (thin horizontal lines) as a function
of $R$ [Eq.~(\protect\ref{eq:R})]. Lines are
guides to the eye.  The thick horizontal lines denote the unperturbed
band edges of the parent o-SL\@.
The vertical
bars on the d-SL data points denote the range of binding energies
obtained from $\sim$10 different realizations of a d-SL with length
$N=128$.}
\label{fig:gaps}
\end{figure}

To understand the possibility of impurity-like localization, consider
for example a $G_{3}$ $\delta$-layer embedded in the otherwise perfect
o-SL $\cdots A_{2}G_{2}A_{2}G_{2}A_{2}G_{2}\cdots$, thus
con\-ver\-ting it in\-to $\cdots
A_{2}G_{2}A_{2}G_{3}A_{2}G_{2}\cdots$.  If the $G_{3}$ $\delta$-layer
is attractive to electrons (holes) it will bind a state below the
conduction-band minimum (CBM) [above the valence-band maximum (VBM)]
of the o-SL \cite{Mader92}.  We find that a (GaAs)$_{3}$
$\delta$-layer in the (AlAs)$_{2}$/(GaAs)$_{2}$ o-SL indeed binds an
electron {\em and\/} a hole [Fig.~\ref{fig:psi}(b)], while an
(AlAs)$_{3}$ layer binds an electron but does not bind a hole
[Fig.~\ref{fig:psi}(c)].  The bound-state localization lengths
$\gamma^{-1}$ and $L_{\text{eff}}$ obtained with a {\em single\/}
$\delta$-layer are similar to those obtained in the fully disordered
SL (Fig.~\ref{fig:psi}), suggesting that the same mechanism of
localization could be at work in both cases.  As one increases the
concentration $R$ of randomly positioned $\delta$-layers, one finds
more bound states which are eventually forming a quasi continuum
inside the band gap \cite{Lifshits63}.
%
% fig. 2
%
This is illustrated in Fig.~\ref{fig:gaps}, where the band-edge
energies of d-SL's with $N =$ 128 ML \cite{length} are plotted as a
function of the degree of disorder $R$.  As $R$ approaches zero, the
band edges merge with the finite binding energies of an isolated
$\delta$-layer [scenario (i)], and not with the unperturbed band edges
of the o-SL [scenario (ii)]. This significant observation suggests
that the localization energy in the d-SL comes mostly from impurity
binding.  Figure \ref{fig:gaps} shows that for large degrees of
disorder $R$, the band edges are pushed further into the gap.  To
isolate the effect of pure disorder from the effect of impurity-like
localization, we also show in Fig.~\ref{fig:gaps} the band edges of an
o-SL containing a {\em periodic array} of $\delta$-layers of
concentration $R$, i.e., separated by a distance $\sim R^{-1}$. We see
that even for an array of closely spaced $\delta$-layers ($R\to1$),
the binding energy does not increase, indicating a negligible
interaction between the neighboring, {\em coherently arranged\/} bound
states.  In contrast, in a d-SL the $\delta$-layer-like bound states
are arranged {\em incoherently\/}, leading to a
band tail.  These studies show that the localization length of the
band-edge states in a d-SL is decided already by the chemical,
impurity-like binding of a single $\delta$-layer
[Figs.~\ref{fig:psi}(b) and \ref{fig:psi}(c)].  Furthermore, the
energy position of the gap levels at small to intermediate degrees of
disorder $R$ is also determined by the properties of non-interacting,
{\em periodically arranged\/} $\delta$-layers
(Fig.~\ref{fig:gaps}). For higher values of $R$, disorder shifts and
broadens the gap levels further into the gap, without modifying their
localization length.

To illustrate the dependence of the gap-level shifts on the particular
form of disorder, we compare the results obtained for the d-SL with
those found in a partially-ordered superlattice (po-SL): Arent et al.\
\cite{Arent94} grew po-SL's with the same distribution function $p(n)$
for the $A_{n}$ layers as in a d-SL, but preserved long-range order by
requiring that each $A_{n}$ layer be followed by a $G_{4-n}$ layer.
Therefore, at positions 1,5,9,\dots there is always an $A$ layer, and
at positions 4,8,12,\dots there is always a $G$ layer, while at the
``sandwiched'' positions, $A$ and $G$ layers are distributed randomly.
We see in Fig.~\ref{fig:gaps} that the presence of long-range order
leads to an even larger shift of gap levels than in the equivalent
d-SL at the same $R$ value.  The band-edge wavefunctions have the same
characteristic $L_{\text{eff}}$ and $\gamma$ as in a d-SL.  Thus,
absence of long-range order (in the d-SL) is not essential for
obtaining large band-edge shifts.

The calculated band-gap reduction of the d-SL and po-SL with respect
to the o-SL (Fig.~\ref{fig:gaps}) is consistent with
experiment \cite{Sasaki89,Arent94}: We find
band gaps of 2.09, 1.94, and 1.87 eV for the o-SL, d-SL, and po-SL,
respectively, compared with the experimental PL emission lines at
2.02, 1.96 \cite{Sasaki89}, and 1.87 eV \cite{Arent94}, respectively.

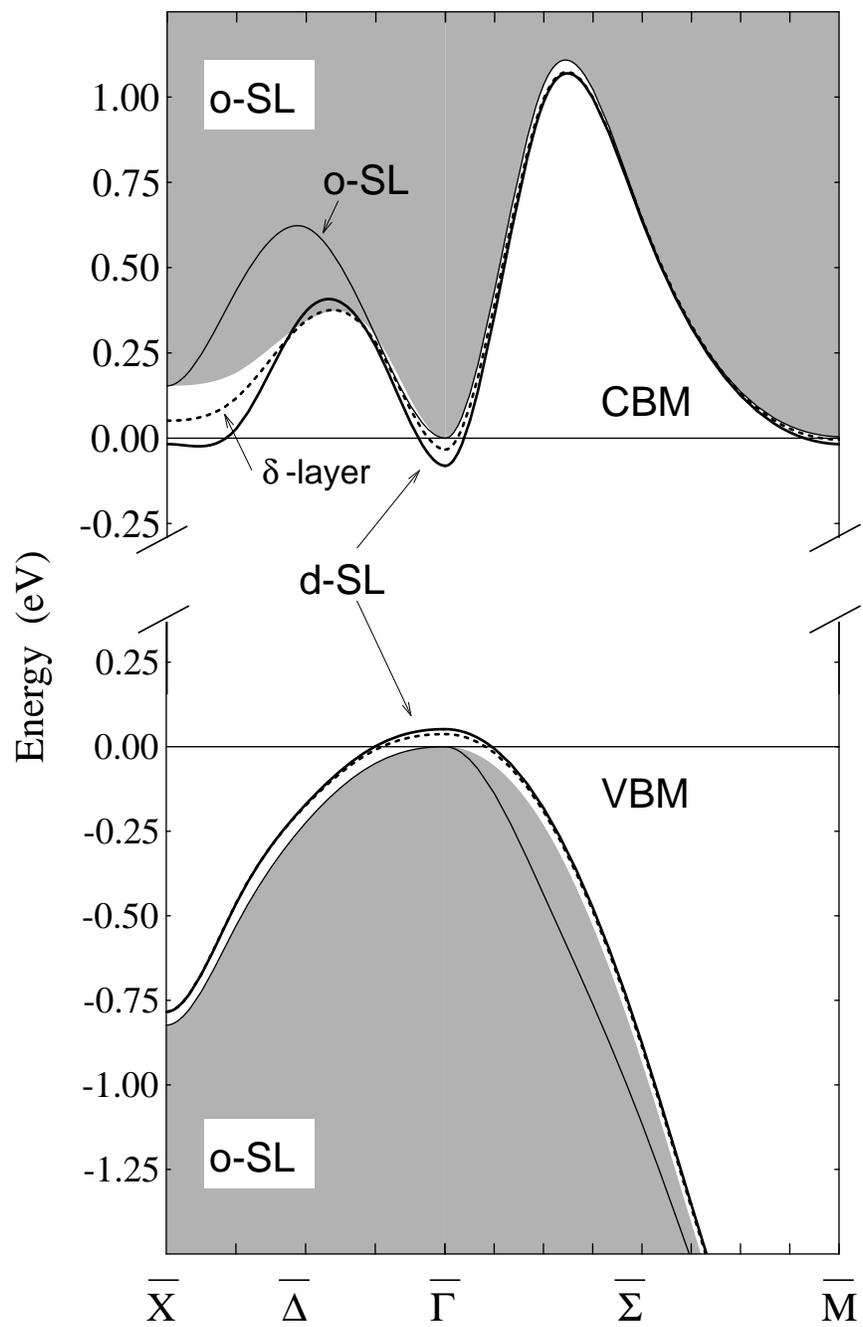
\begin{figure}
%\hbox to \hsize%
%{\epsfysize=1.00\hsize\hfil\epsfbox{figures/dsl-dispersion.eps}\hfil}
%\nobreak\bigskip
\caption{Dispersion of the band-edge state in a d-SL (solid lines),
o-SL (thin lines), and a single $\delta$-layer in an o-SL host (dotted
lines) along the $\bar\Delta$ and $\bar\Sigma$ symmetry lines in the
planar Brillouin zone. Also shown are the (001)-projected bands of the
o-SL (shaded area). The supercell size was $N=128$.}
\label{fig:disp}
\end{figure}

To determine if the d-SL has a direct or indirect gap in the
two-dimensional Brillouin zone, we show in Figure~\ref{fig:disp} the
dispersion of the band-edge states of the d-SL (solid lines), o-SL
(thin lines) and single $\delta$-layer (dotted lines) along the
symmetry lines $\bar\Sigma$ and $\bar\Delta$, i.e., from $\bar\Gamma$
to $\bar M = \frac{1}{\sqrt{2}} (1,1)$ and from $\bar\Gamma$ to $\bar
X = \frac{1}{\sqrt{2}} (1,0)$, respectively.  The thin horizontal
lines denote the band edges of the underlying o-SL.  We find that the
conduction bands of the d-SL dip below these lines.  The difference
(``binding energy'') increases in the order $\bar M \to \bar\Gamma \to
\bar X$. \cite{L-repulsion}
Thus, the large binding energy at $\bar\Gamma$ pulls the
conduction-band edge below the one at $\bar M$ by 60 meV, {\em making
the d-SL a direct-gap material\/} \cite{tb}, even though the o-SL is
indirect (with CBM at $\bar M$).
%
% fig. 3
%
This suggests strongly that the observed \cite{Sasaki89}
strong PL intensity is due to the occurrence of a direct band gap,
leading to efficient recombination of electrons and holes localized in
the same spatial region along $z$ [see, e.g., the same positions along
the chain of states v2 and c3 in Fig.~\ref{fig:psi}(a)]. The
localization along $z$ relaxes the $\bbox{k}_{\parallel}$-selection
rule, thus further enhancing the oscillator strength.  The enhanced
oscillator strength is reflected by short radiative lifetimes $\tau$:
we calculate $\tau =$ 1 ns for the v2$\to$c3 transition at energy 1.96
eV [Fig.~\ref{fig:psi}(a)].  These radiative lifetimes are
1000$\times$ faster compared to those measured in indirect-gap o-SL's
($\tau \approx 5.5$ $\mu$s at $T =$ 2 K) \cite{Ge94}.

%
% fig. 4
%

To understand why the PL intensity in a d-SL has a weaker temperature
dependence than in an o-SL, consider vertical transport to
non-radiative recombination centers. This channel will be inhibited,
unless there are strongly overlapping electron states within an energy
$\sim k T$ of each other. Strong overlap occurs when $\zeta(E,T) =
k T \rho(E_{T}) L_{\rm eff}(E_{T}) > 1$, where $\rho(E)$ is the
one-dimensional density of states, and $E_{T}$ is a typical energy
that is thermally occupied at temperature $T$.  Figure~\ref{fig:DOS}
shows $\rho(E)$, $L_{\rm eff}(E)$, and $\zeta(E,T)$ obtained by
averaging over 100 realizations of $N=2000$ ML d-SL's, calculated
within the Kronig-Penney approximation with band offset fitted to our
pseudopotential calculations.  We find that: (i) the DOS exhibits a
$\sim$200 meV wide band tail extending below the CBM of the o-SL (the
zero of energy); (ii) $L_{\rm eff}$ is almost constant ($\sim$20 ML)
in the band tail; (iii) at the thermally populated levels [denoted by
the arrows in Fig.~\ref{fig:DOS}(b)] $\zeta \ll 1$, thus vertical
hopping is suppressed.  Consequently, the d-SL will have a weaker
temperature dependence of the PL decay, because higher temperatures
will be needed to dissociate the electron-hole pairs in the vertical
dimension.

Acknowledgment---Fruitful discussions with D. J. Arent and S.-H.\ Wei
are gratefully acknowledged. This work was supported by the Office of
Energy Research, Materials Science Division, U.S.\ Department of
Energy, under grant No.\ DE-AC36-83CH10093.

% for galley:
%\vskip -12 pt
%
% REFERENCES references %%%%%%%%%%%%%%%%%%%%%%%%%%%%%%%%%%%%%%%%%%%%%
%

% end here for galley proofs
%\end{document}
%\endinput
%\clearpage

\narrowtext

\narrowtext

\narrowtext

\narrowtext
\begin{figure}
%\hbox to \hsize%
%{\epsfxsize=0.90\hsize\hfil\epsfbox{figures/dsl-dispersion.eps}\hfil}
%\nobreak\bigskip
\caption{(a) Density of states $\rho(E)$ (solid line),
$L_{\rm eff}(E)$ (dashed line) averaged over 100 realizations of
$N = 2000$ ML d-SL's, calculated with the Kronig-Penney model.
Dots indicate
$L_{\rm eff}(E)$ obtained for an $N = 128$ ML
pseudopotential calculation.  (b) The product $\zeta(E,T) = k T \rho(E)
L_{\rm eff}(E)$.
}
\label{fig:DOS}
\end{figure}


\begin{references}
%\bibitem{SL}\vskip -1.0cm J. L. Beeby et al.\ (editors), {\em
%Condensed Systems of Low Dimensionality}, NATO ASI Series B {\bf 253},
%(Plenum Press, New York, 1991).
\bibitem{SL} J. L. Beeby et al.\
(editors), {\em Condensed Systems of Low Dimensionality}, NATO ASI
Series B {\bf 253}, (Plenum Press, New York, 1991).
\bibitem{Chomette86} A. Chomette et al., Phys.\ Rev.\ Lett.\ {\bf 57},
1464 (1986).
\bibitem{Sasaki89} A. Sasaki et al., Jpn.\ J. Appl.\
Phys.\ {\bf 28}, L1249 (1989); J. Crystal Growth {\bf 115}, 490
(1991).
\bibitem{Borland63} R. E. Borland, Proc.\ R.\ Soc.\ A {\bf
274}, 529 (1963).
\bibitem{Papa76} C. Papatriantafillou and
E. N. Economou, Phys.\ Rev.\ B {\bf 13}, 920 (1976).
\bibitem{Pavesi89} L. Pavesi et al., Phys.\ Rev.\ B {\bf 39}, 7788
(1989).
\bibitem{EGWang94} E. G. Wang et al., Appl.\ Phys.\ Lett.\
{\bf 64}, 443 (1994).
\bibitem{Mader94} K. A. M\"ader and A. Zunger,
Phys.\ Rev.\ B, BT5032, in press (1994).
\bibitem{Wang94} L.-W.\ Wang
and A. Zunger, J. Chem.\ Phys.\ {\bf 100}, 2394 (1994).
\bibitem{Mader92} K. A. M\"ader and A. Baldereschi, Mat.\ Res.\ Soc.\
Symp.\ Proc.\ {\bf 240}, 597 (1992).
\bibitem{Lifshits63}
I. M. Lifshits, Sov.\ Phys.\ JETP {\bf 17}, 1159 (1963).
%\bibitem{Lifshits87} I. M. Lifshits, S. A. Gredeskul, and
%L. A. Pastur, {\em Introduction to the theory of disordered systems},
%(Wiley and Sons, New York, 1987), p. 62.
\bibitem{length} Note that
in order to correctly describe localized wavefunctions we need $N \gg
L_{\text{eff}}$.  We have confirmed that band-edge wavefunctions
calculated in an $N=128$ d-SL agree very well with those shown in
Fig.~\protect\ref{fig:psi}(a), which were calculated with $N=1000$.
\bibitem{Arent94} D. J. Arent et al., Phys.\ Rev.\ B {\bf 49}, 11173
(1994).
\bibitem{L-repulsion} The large binding energy at $\bar X$ is a
consequence of the level repulsion of the folded $L_{\rm 1c}$ states,
which is much stronger for odd values of the repeat period $n$ than
for even $n$ [See, for example, S.-H.\ Wei and A. Zunger, Appl.\ Phys.\
Lett.\ {\bf 53}, 2077 (1988)].
In the d-SL the odd-even selection rule is broken,
leading to a stronger level repulsion in the d-SL than in the $n=2$ o-SL.
\bibitem{tb} In a recent tight-binding study of an $N = $
10--20 ML model of a d-SL, Wang et al.\ \protect\cite{EGWang94} found
a nearly dispersionless conduction band along $\bar\Delta$ and
$\bar\Sigma$, an {\em indirect gap\/} at $\bar M$, and a 2--4 ML
localization length.  The differences with respect to the present
pseudopotential calculation may reflect a combination of the use of
rather short SL's and the restricted variational flexibility of the
tight-binding method used in Ref.~\protect\cite{EGWang94}.
\bibitem{Ge94} W. Ge et al., J. Luminesc.\ {\bf 59}, 163 (1994).
\end{references}
\end{document}